\documentstyle[12pt]{article}
\begin{document}
\newcommand{\newc}{\newcommand}
\title{NEW TRENDS IN PARTICLE
THEORY \thanks{ Plenary talk at the  EPS Marseilles Conference, July 1994}}

\author{Luis Ib\'a\~nez,\\ Departamento de
Fisica Te\'orica \\                       Universidad Aut\'onoma de Madrid
\\        Cantoblanco, 28034 Madrid\\
\\
}
\maketitle
 \vspace{-9cm}
\hfill FTUAM-4/94
\vspace{14cm}
\begin{abstract}
I discuss some  new trends in Particle
Theory beyond the Standard Model. Some topics which are briefly
covered include electroweak baryogenesis, gauge versus non-gauge discrete
 symmetries,
the strong-CP problem, dynamical symmetry breaking scenarios,
supersymmetric grand unification, new aspects in low-energy
supersymmetry and superstring phenomenology.

\end{abstract}
\maketitle

\newpage

\vskip 5mm
\section{ Introduction}
\vskip 2.5mm
The official title of this talk is probably too ambitious. I guess nobody knows
 what are the
future trends of Particle Theory. Instead of that
I will just review some work done in the general field
of Physics Beyond the Standard Model in the last couple of years or so. I will
 skip here the
standard praising of the Standard Model (SM) and the immediate list
of reasons why there should be something else Beyond the Standard Model
(BSM). Instead of that let me desplay for you the present situation of the
Stock Market of BSM ideas:
\bigskip

- Susy phenomenology {$\Uparrow $}

- Weak-scale baryogenesis {$\Uparrow $}

- Astroparticle (solar $\nu $'s, COBE...)  {$\Uparrow $ }

- String phenomenology  {$\Uparrow $ }

- $W_L-W_L$ scattering at LHC/SSC {$ \Uparrow $}

- Constraints on BSM from EW-loops  {$  \Uparrow $}

- Non-commutative geometry models    {$ \uparrow $ }

- Technicolor and ETC   {$ \updownarrow $ }

- Axions; Global symmetries   {$ \updownarrow $ }

- B-violation at high {\it energies} in SM   {$ \downarrow $}

- $t-{\bar t}$ condensates  {$ \downarrow  $ }

- Wormholes solving the Cosm.C. problem  {$ \Downarrow $ }

- Non-SUSY $SU(5)$  {$ \Downarrow $ }

- 17 KeV neutrino {$  \Downarrow \Downarrow $ }

\bigskip
As in real life, this is a very speculative market and the
declared tendency of each particular topic is not directly
related to the intrinsic value of each idea and is probably
even less directly connected to reality. Of course, there are
many interesting topics which are not in the list and even
those in the list cannot be througly reviewed. Fortunately,
some of the most interesting topics have been discussed by
other speakers in this conference: Astroparticle physics
(Silk), neutrino physics (Spiro, Petcov,..), constraints on
BSM physics from electroweak loops (Altarelli...), extra
$Z$-bosons (Cvetic, Taxil) etc. Concerning neutrino physics
I just would like to make a very trivial comment: we should
not overemphasize neutrino mass estimates based on the see-saw
mechanism(s). These type of estimates are purely {\it qualitative}
and should only be taken as such. Let us now turn to review a few
topics which have received much attention in the last couple of	years
or so.
\vskip 2.0mm

\section{  B and L-number violation in the SM at high temperatures}
\vskip 2.0mm
The SM has baryon number $B$ and the three lepton numbers $L_i$ as
accidental global $U(1)$ symmetries. They are good classical
symmetries of the {\it minimal } SM but are violated by quantum
effects associated to the electroweak interactions. Indeed, the
currents associated to $B$ and $L_i$ in the SM have mixed anomalies
with the $SU(2)_W$ gauge bosons, and hence these symmetries are not
respected  by quantum mechanical effects. It was 't Hooft [1] who
first realized that $SU(2)_W$ non-perturbative (instanton) effects can
give rise to $B$ and $L_i$-violating effective interactions.
Numerically, the rate of these interactions is of order
$|T|^2\simeq e^{-S}\simeq e^{-4\pi / {\alpha _W}}\simeq 10^{-80}$,
and hence negligible for all practical purposes. However, it turns
out that at {\it high temperatures} the $B$-violating interactions are no
longer supressed [2]. There are classical Higgs and $SU(2)_W$ gauge-field
configurations (''sphalerons"[3]) which interpolate between vacua
with different $B$-(and $L$-) number. Thus at high temperatures
($T \geq M_W/\alpha _2$) $B,L$-violating interactions are unsupressed.
The above fact has two important (cosmological) consequences:

\vskip 2.0mm
i) {\it Weak-scale baryogenesis}
\vskip 2.0mm
It could well be that these electroweak effects could be the source
of $B$-violation required to generate the primordial baryon/antibaryon
asymmetry of the universe[4]. However, the other two ingredients [5] required
for this generation, CP-violation and departure from thermal equilibrium,
do not seem so easy to get within the minimal SM. Although the SM
automatically has a source of CP-violation from the KM-phase, this turns
out to be too small (for recent controversy about this point see
ref.[6]). This can
easily be cured by a modest extension of the SM including extra scalars
(or by going to the SUSY-SM which has additional sources of phases).
The third point, breakdown of thermal equilibrium, is the toughest to get.
Here the nature of the electroweak phase transition is of the outmost
importance. It turns out that, contrary to what one would naively expect,
the character of this very fundamental phase-transition is poorly
known. If the phase transition is first order, baryogenesis is feasible.
 Estimations suggest that one should not expect a first order phase
transition in the minimal SM for Higgs masses $m_H\geq 60$ GeV. Taking
into account the LEP data, this leaves little space for a first order
phase transition to develop. There are, however many theoretical
uncertainties involved
(see Fodor's contribution to these proceedings)
and one cannot rule out completely this possibility.
On the other hand, extending the SM by adding a non-minimal Higgs sector
 makes life easier concerning baryogenesis.
\newpage
ii) {\it Erasure of a primordial baryon asymmetry}
\vskip 2.0mm
It could well be that the low-temperature baryogenesis scenario outlined
above does not work, in which case one could look back to the more
traditional high-temperature baryogenesis scenarios which were so 	popular
in the eighties [7]. Those schemes take the $B$ and  CP-violation
from explicit couplings present in most Grand Unified Theories (GUTs)
like $SU(5)$ or $SO(10)$.
Departure from thermal equilibrium is in this case very easy to obtain
from late decay of superheavy particles (e.g. coloured scalars)
generically present in GUTs. These scenarios are not free of problems
either: inflation may completely dilute any primordial baryon asymmetry
created in this way unless the reheating takes place at temperatures
$\leq  10^{13}$ GeV. In the last few years it has also been realized
that the high temperature electroweak effects discussed above may also
erase any primordial $B$ asymmetry, since those effects are in thermal
equilibrium below temperatures $T\leq 10^{12}$ GeV or so. There is a
loop-hole though in this argumentation. If a net $B-L$ density is
generated at a primordial stage, electroweak effects will be unable
to erase it. This is because the combination $B-L$ has $no$ mixed anomaly with
$SU(2)_W$, and hence this symmetry is respected by all standard model
interactions [2]. Thus the idea is to generate a primordial $B-L$ density
in some GUT scenario like $SO(10)$ (minimal $SU(5)$ does not work because it
has an exact global  $B-L$ symmetry). High temperature electroweak
effects may partially convert a baryon asymmetry into a lepton
asymmetry (or viceversa) but are unable to erase the net $B-L$ [8].

If one takes this option of a primordial $B-L$ generation, one may
still be in trouble if there are additional explicit interactions in
the effective Lagrangian violating $B$ and/or $L$. Even if these
additional interactions are very tiny or even supressed by inverse powers
of large masses (like GUT scale or Planck mass), they may be
sufficiently effective in erasing the primordial asymmetry. If one insists
in preserving this asymmetry one can obtain constraints on $B$ and/or
$L$-violating terms like e.g., Majorana  masses for the left-handed
neutrinos. One finds for any of the three neutrinos the bound [9]
\begin{equation}
m_{\nu _i}\ \leq \ 10^{-3}\ eV
\label{eq:l}
\end{equation}
if the primordial B-asymmetry is to be preserved. Other stringent limits
are also found for other $B$ and/or $L$-violating terms [10] present in
different models (e.g. $R_p$-violating operators in the SUSY-SM). If the
above limits are strict, this would mean bad news for $\nu $-oscilation
experiments (or high-temperature baryogenesis!). More recently it has
been realized that there are different effects which may somewhat
relax these limits:

 i) In the presence of supersymmetry, due to the
existence of a new aproximate global $U(1)$ symmetry (R-symmetry)
beyond $B$ and $L$, electroweak effects are again unable to erase
a primordial $B$-asymmetry [11]. Since SUSY is not an exact symmetry, its
only effect is to postpone the electroweak erasing down to temperatures
$T\leq 10^7$ GeV or so (instead of $T\leq 10^{12}$ GeV). This is
sufficient to relax e.g. the $\nu $-mass bounds to
$m_{\nu }\leq 10 $ eV. This allows for more brilliant prospects for
neutrino oscillations.

ii) Fermion mass effects on the baryon number densities may also be
important. It has  recently been pointed out that those
may also avoid the erasure of a primordial asymmetry (see Dreiner's
contribution to these proceedings).

iii) The smallness of the electron Yukawa coupling makes that the $e_R$
gets very late into thermal equilibrium. This effect turns out to relax
also the above type of bounds drastically [80].

Both in the high- and low-temperature scenarios for baryogenesis,
it is clear that one cannot neglect the $B/L$-violating electroweak
effects. Imposing that a given BSM scheme is consistent with either
one or the other of these scenarios may be a  extremely effective
constraint on physics beyond the standard model.
For example, minimal $SU(5)$ cannot yield any baryon asymmetry
since it will  necessarily be diluted by electroweak effects.
Other mechanisms
to generate baryon asymmetry may be at work, one of the most
interesting ones being the one in ref.[12].

A final speculative remark is in order. I mentioned above that
$B-L$ is an anomaly-free global symmetry. This is only partially
correct: it does not have mixed anomalies with SM interactions but
it {\it does } have mixed {\it gravitational} anomalies. Thus one
expect the existence of gravitational effects violating $B-L$.
One could speculate [13] on the possibility that at extremely high
temperatures,
just below the Planck mass, these gravitational effects could
generate a $B-L$ primordial asymmetry. The usual electroweak
effects would just redistribute the relative ammounts of $B$ and $L$.
This scenario would have the beauty that the very existence of
an asymmetry would be a direct consequence of the anomaly
structure of the SM. But, although atractive in principle,
it is not obvious how
to make such an scenario to work out in detail.

\section{ The Crisis of Global Symmetries}
\vskip 2.0mm
We mentioned above the four  global $U(1)$ symmetries of the SM
($B$ and $L_i$). They are {\it accidental} symmetries of the theory
and are a mere consequence of $SU(3)\times SU(2)\times U(1)$
gauge invariance, Lorentz invariance and renormalizability. In
going to physics beyond the standard model one normally
needs to impose new global symmetries to achieve different
phenomenological goals. Examples of those are: 1) The Peccei-Quinn
$U(1)_{PQ}$ symmetry, introduced to solve the strong-CP problem ;
2) The discrete global symmetries introduced in multi-Higgs
models in order to avoid flavour-changing neutral currents (FCNC);
3) Discrete (or continuous) ''horizontal" symmetries introduced
in order to get appropriate ''textures" for the fermion mass
matrices ; 4) The usual $Z_2$ R-parity (or other ''generalized
matter parities") imposed in the minimal supersymmetric
standard model (MSSM) in order to avoid fast proton decay.

All these global symmetries are really imposed by hand and,
unlike gauge symmetries, they are not motivated by  any
{\it fundamental} symmetry principle. In the last few years
it has been realized that a number of gravitational effects
badly violate {\it global} symmetries: terms which are forbidden
from a Lagrangian by impossing a global symmetry (continuous or
discrete) are regenerated by gravitational dynamics (wormholes,
blackholes [14]). Thus {\it global symmetries are not effective}
in fulfilling their expected duties! This is what I call in this
section {\it The Crisis of Global Symmetries}.

A good example of the effect of this ''crisis" is the
difficulties which are expected for a Peccei-Quinn type of
solution for the strong-CP problem. This solution [15] requires the
existence of a {\it global} $U(1)_{PQ}$ symmetry which has
mixed anomalies with QCD. This symmetry is spontaneously
broken, giving rise to the corresponding (pseudo-)Goldstone
boson, the {\it axion}. Due to the anomaly, the axion field
{\it a} behaves as an ''effective $\theta $ parameter" which
couples to gluons as  $a/f_aG^{\mu \nu }{\tilde G}_{\mu \nu}$.
Non-perturbative QCD effects generate a scalar potential for {\it a}
of the form
\begin{equation}
V_{QCD}(a)\ =\ (m_a^{QCD})^2f_a^2 \ (1\ -\ cos{\bar a})
\label{eq:2}
\end{equation}
where ${\bar a}=a+\theta $. This potential is minimized for,
$<{\bar a}>=0$, yielding an elegant solution to the strong CP-problem
(the problem of the unexpected smallness of the QCD $\theta $
parameter). Let us now assume that there are new operators
of dimension $D=4+n$ which explicitly violate the $U(1)_{PQ}$
global symmetry. As I discussed above, gravitational effects are
expected to generate such terms. In this case there will be additional
contributions to the axion scalar potential. A simple estimation 	leads
to a contribution [16]
 \begin{equation}
V_{grav}(a)\ =\ (m_a^{grav})^2f_a^2 \ (1\ -\ cos(n{\bar a}\ +\ \delta ))
\label{eq:3}
\end{equation}
where $\delta $ is a number of order one. This potential is no longer
minimized for $<{\bar a}>=0$ and hence, in order not to spoil the solution
provided by eq.[2], $m_a^{grav}$ has to be very small. In order to keep
a $\theta $ parameter $\leq 10^{-10}$, as required by experiment, one needs to
 have
 \begin{equation}
(m_a^{grav})^2\ \leq \ 10^{-10}\ (m_a^{QCD})^2
\label{eq:4}
\end{equation}
A naive estimation tells us that in order for this contribution to be so
supressed, the operator of lowest dimension $D$ violating $U(1)_{PQ}$
needs to have $D\geq 12$!! Thus an apropriate (gauge) symmetry should thus
guarantee the absence of operators with $D\leq 12$ for the
Peccei-Quinn mechanism to work. This looks like a bit hard to get.

There are in the literature other proposed non-axionic solutions to the
strong-CP problem:

i) Assume that the laws of nature are CP conserving ($\theta _{bare}=0$)
and that this symmetry is spontaneously broken [17]. Once this symmetry
breaking
occurrs, a non-vanishing (calculable) $\theta $ will appear and , hence
one has to cook carefully the model in order to get small loop corrections
to $\theta $. Although this is certainly a possibility, the actual
realizations of this general idea are quite contrived.

ii) If the ''current" u-quark mass vanishes, there is a global
chiral $U(1)$ symmetry which allows to ''rotate away" the
$\theta $ angle. This possibility [18] is very neat and simple but
 hard to reconcile with the standard lore of chiral Lagrangian
estimations of light-quark masses. Given the simplicity
of this possibility I think it should however be seriously
reconsidered.

Notice that these two alternatives also need the existence of
global symmetries (CP in the first case, a chiral $U(1)$ in the
second). Thus these two alternatives are also jeopardized by
generic gravitational effects!

The case of the proposed solutions
to the strong-CP problem is just an example. There are other
BSM schemes which make use of discrete symmetries in a
fundamental way and are also in trouble. A second prominent
example is the R-parity discrete symmetry of the
minimal supersymmetric standard model (MSSM). Unlike what happens in the
simple standard model, in its  SUSY version $B$ and $L_i$ are {\it not}
automatic accidental symmetries of the theory. In particular, the most
general Lagrangian consistent with SUSY, $SU(3)\times SU(2)\times U(1)$
and Lorentz invariance allows for Yukawa couplings which violate
$B$ and/or $L_i$ symmetries. This would be phenomenologically
catastrophic and hence one imposses by hand some {\it global} discrete symmetry
which forbids the dangerous couplings. The simplest example of such a
symmetry is ''R-parity", a $Z_2$ symmetry under which the
usual SM particles are even and all their SUSY-partners are odd.
This symmetry forbids all $B$ and $L_i$-violating couplings.
In view of our previous discussion, a global symmetry is not
enough protection and the MSSM would be in trouble.

\vskip 5.4mm
\section{ A New Guiding Principle?}
\vskip 2.0mm
There is a way out to the above ''global symmetry crisis". If the
symmetries which are phenomenologically required are
{\it gauge symmetries}, they will be immune to the
problematic gravitational effects. It is well known that
those effects cannot e.g. violate charge conservation which is
a symmetry asociated to a gauge theory (QED).
The intuitive reason for
this difference between global and local is simple. The most
important characteristic of {\it local} symmetries is that
they play an important role in fixing what are the {\it actual
physical degrees of freedom} of a theory. Local symmetries
not only tell us what terms are allowed or forbidden in the
effective Lagrangian (this is also done by the global
symmetries) they also allow us  to get rid of spurious
non-physical states in the theory. Gravitational (or any other)
perturbative or non-perturbative effects may generate terms
violating a global symmetry but they can never modify the
number of physical degrees of freedom of a theory and, hence, they
cannot violate a gauge symmetry.
This suggest to impose the
following physical principle:

{\it ''All symmetries (even discrete ones)
should be
{\bf gauge} symmetries
(unless they are accidental)"}

By accidental symmetries I mean symmetries like baryon and
lepton numbers, which are a mere consequence of renormalizability
and gauge invariance. This could be a good possibility to obtain
aproximate Peccei-Quinn $U(1)_{PQ}$ symmetries in specific models [19].
Notice that Peccei-Quinn symmetries are anomalous and, hence,
cannot be gauged in a straightforward way. It is interesting to remark, though
that in string models PQ symmetries may be gauged
under certain conditions [20].

The above principle tells us that even {\it discrete symmetries should be
gauged}. Many particle theorists are not familiar with the concept of
discrete gauge symmetry. The most intuitive way to generate a
discrete gauge symmetry (DGS) is to start with a standard gauged
$U(1)$ theory and break that symmetry spontaneously through the
vev of a scalar field [21]. If the scalar field has charge $Nq$ and the
rest of the particle spectrum has charges $q_i=M_iq$, with not all
$M_i$ equal to a multiple of $N$, one can check that there is an
unbroken $Z_N$ subgroup of the original $U(1)$ which remains unbroken.
Locally, there is no
way to distinguish a gauged from a global discrete symmetry but gauge
symmetries give rise to some non-local interaction effects of the
Bohm-Aharanov type which are not present in the global symmetry case [22].
These effects may have only, at most, cosmological relevance. However,
there are other two important practical differences compared to
the global case. One of them we already mention, DGS are immune to
destabilizing gravitational effects. The other is that, just like it
happens with gauged $U(1)$ symmetries, {\it discrete symmetries
should be anomaly-free} [23]. The discrete charges of chiral fermions
should obey certain restrictive discrete anomaly
cancellation conditions. These conditions look very much like
discretized versions of usual anomaly cancellation conditions. For
example, the mixed $Z_N$-$SU(M)$ anomaly cancellation conditions
look like [23]
 \begin{equation}
\sum _i(q_i)\ =\ 0\ \ mod\ \ N
\label{eq:5}
\end{equation}
where $exp(i2\pi q_i/N)$ is the $Z_N$ charge of each of the $SU(M)$
fermion $M$-plets.

The discrete anomaly cancellation conditions should be obeyed by
any DGS and this may lead to interesting phenomenological
implications. For example the discrete $Z_N$ symmetries guaranteing
sufficient proton stability in the SUSY standard model
(''generalized matter parities") were clasified in ref.[23].  It was found
that only four of them (a $Z_2$ and three $Z_3$s) are discrete
anomaly-free with the particle content of the MSSM. The usual R-parity is
one of them. Of course, this anomaly-freedom criterium may be
applied to many other BSM schemes like, for example, discrete symmetries
giving rise to appropriate quark-mass matrix textures.

It is also worth remarking that string theory does not have much sympathy
for global symmetries either. In fact, there is a general theorem in strings
 which states
that any exact continuous (e.g., $U(1)$)  symmetry cannot be just global, it
has
 to be a gauge
symmetry [24]. An equivalent theorem for discrete (e.g. $Z_N$) symmetries has
 not
been proved. However, it has been shown  in explicit four-dimensional
strings that many of the $Z_N$ symmetries have a gauge origin, and that could
well be the case for all discrete symmetries in string models. In particular,
it has been shown that CP itself may be understood as a discrete gauge symmetry
 [25].
It has also been shown in plenty of four-dimensional string examples
that the discrete anomaly cancellation conditions mentioned above are indeed
satisfied [26]. The origin of this cancellation is still unclear but, like in
 the
continuous gauge case, it is probably a consequence of the important
''modular invariance" property of string theories.

\vskip 5.4mm
\section{ The Naturality Problem: the Strongly Interacting Approach}
\vskip 2.0mm
By ''naturality problem" I mean the problem of the instability of the Higgs
sector of the standard model under quantum corrections. This problem goes under
different names in the literature: gauge hierarchy problem, the mass problem
 etc.
This is a problem which has concerned many particle theorists since  more than
fifteen years ago. Although a minority of physicists still mantain that this is
not a problem since one can always renormalize the scalar mass to the value we
wish, the immense majority think that there is indeed a problem since using
physical (cut-off) regulators we need to make ridiculous fine-tunings to
mantain the Higgs scalar sufficiently light to really induce $SU(2)\times U(1)$
breaking. As is well known, there are still essentially two schools of thought
concerning this problem: i) The strongly interaction scenarios
and ii) the supersymmetry approach. Let me say a few words about the first of
 these
and postpone the second to the next section.

 The strongly interacting schemes assume that, at energies not much above
the weak scale, there are new {\it strongly interacting} phenomena which will
allow us eventually to understand the origin of the symmetry-breaking (Higgs)
sector of the standard model. In these schemes the Higgs field  (and sometimes
also the quarks and leptons) are {\it composite} particles.
The main problem of this approach a priori is that very little is
known about the non-perturbative physics of {\it chiral gauge theories}. Hence
what one normally does is to imitate the physics of the only relativelly well
understood non-perturbative gauge theory, QCD, which is not chiral. This is
what
 is
done in the simplest and most attractive scenario of this type, Technicolor.

In Technicolor [27] one assumes that there are new QCD-like strong interactions
 at a scale
of order a TeV with gauge group $G_T$. The theory contains fermions coupling to
$G_T$ called technifermions $\Psi _i$. Instead of a Higgs vev, one assumes that
there are non-vanishing vevs for technifermion bilinears, $<{\bar {\Psi
}}_R\Psi
 _L>
\not= 0$, breaking $SU(2)\times U(1)$ spontaneously. All this works very nicely
 and
the $W$ and $Z^0$ bosons indeed get a mass in the usual way. However, and this
 is the
key problem, at this level the quarks and leptons remain massless. In order to
 provide
masses to fermions, the best idea available is the introduction of extra new
 gauge
interactions, {\it Extended Technicolor (ETC)} [28], whose crucial property is
 that they
connect the usual quarks and leptons with the technifermions. The latter are
 massive,
they get a dynamical mass due to technicolor interactions. Then one can draw
 one-loop
graphs in which a quark (lepton) splits into a techniquark and an ETC boson
 which
then recombine again into a quark (lepton). These loops provide masses for the
quarks and leptons of order
 \begin{equation}
m_{q,l}\ \simeq \ {{g_{ETC}^2}\over {(4\pi ^2)}}
{{<{\bar {\Psi }}_R\Psi _L>}\over {M_{ETC}^2}}
\label{eq:6}
\end{equation}
where $g_{ETC}$ is the gauge coupling and $M_{ETC}$  the mass scale of
the new gauge ETC interactions. The idea of ETC is
very nice in principle but very problematic in practice. The above formula has
 to
provide masses for all quarks and leptons. It is very hard for such a one-loop
 effect
to generate masses big enough to account for the masses of
 the third generation of quarks and leptons (particularly
the t-quark). To increase the above contributions one has to either increase
the
value of the condensate $<{\bar {\Psi }} _R\Psi _L>$ or to decrease the ETC
gauge boson masses $M_{ETC}$.  To increase the condensate without increasing
at the same time the $W$ and $Z^0$ masses is not easy. The second
possibility is also very problematic: the ETC gauge bosons necessarily change
 flavour leading to
enormous FCNC unless $M_{ETC}\geq 100$ TeV or so. There is an extra problem for
 the ETC
theories. Usually these theories not only produce dynamically the Goldstone
 bosons
required for the $W$s and $Z^0$ to get massive, they also give rise to a
plethora of other composite scalars ({\it pseudo-Goldstone bosons}), some of
which should have already been seen at present accelarators like LEP.  All
these
(and other) problems
lead to a decline in the popularity of the Technicolor and ETC ideas during the
years 1982-1988 or so.

There has been in the last few years a certain ''discrete revival" of
 Technicolor
and of the strongly-
interacting Higgs sector schemes in general [29]. Concerning ETC,
work has been done in trying to avoid the FCNC problem of these theories. In
 this
connection, two main lines have been explored. The idea in both schemes is
 enhancing the value
of the condensate $<{\bar \Psi }_R\Psi _L>$ without increasing the values of
the
masses of the $W$ and $Z^0$, which are fixed by the ''technipion" decay
constant
$F_{T\pi}$. One of the ideas to achieve this goes under the name of
{\it walking technicolor} [30]. It was pointed out that, if the $\beta
 $-function of the
Technicolor interactions is small, there is an enhancement of the ratio
$<{\bar \Psi }_R\Psi _L>/F_{T\pi}^3$. It turns out however that the achieved
enhancement is not enough to account for the mass of the third generation
 fermions.
The second main idea put forward to increase this ratio is the use of
{\it ultraviolet fixed-point models} [31]. The idea is that, if Technicolor
 interactions
have an ultraviolet fixed point (a zero of the $\beta (\alpha )$-function) at a
 finite
value of $\alpha $, the ${\bar \Psi }_R\Psi _L$ bilinear has large ''anomalous
dimensions", i.e. it gets a large enhancement factor. The problem is that it is
 not
clear whether examples of non-Abelian theories of the above characteristics
 exist.
It has been recently pointed out that non-Abelian gauge theories with a large
number of fermions could have this property. Assuming that a Technicolor theory
with the required properties exist, one can do some interesting
model-building [32].

In spite of the  efforts, no completely compelling ETC model with
 phenomenological
promise has been constructed up to now. Furthermore, as I mentioned above, very
special properties have to be assumed for a Technicolor theory to have any
chance of surviving. On the other hand, these difficulties may well be
due to our lack of understanding of non-perturbative gauge theories, and not
 really
to the idea itself.

An alternative to Technicolor which has received attention in the recent past
is based on the assumption of top-antitop quark condensation, as a substitute
 for
techniquark condensation [33]. This is suggested by the fact that the top-quark
 is
extraordinarily heavy compared to the other quarks. It is assumed that a
bilinear condensate $<{\bar t}_Rt_L>\not= 0$ forms due to some unknown
strong interactions giving rise to $SU(2)\times U(1)$ breaking. In the original
formulation of this idea, these unknown interactions were described by a
 Nambu-Jona Lasinio
type of model which lead to some interesting predictions like
 $m_{Higgs}=2m_{top}$ and
some more problematic results like $m_{top}\geq 210$ GeV. This latter result
 gives
rise to large loop contributions to the $\rho $-parameter which are several
 standard
deviations away from the electroweak data. There are also extra theoretical
 concerns:
nothing is known about the origin of the masses of the rest of the quarks and
 leptons
nor about  the origin of the top condensation. It seems to me that the general
idea of the t-quark condensates is much better that the actual implementations
done up to date.

Due to our lack of knowledge of the precise strongly interacting dynamics which
 could
be waiting for us above the weak scale, perhaps the wiser approach is trying to
parametrize the process of $SU(2)\times U(1)$  breaking in the most general
possible way. In this connection, the most promising way seems to use the
effective (chiral)-Lagrangian approach which is used so succesfully
in describing the chiral symmetry-breaking dynamics of QCD [34]. The effective
chiral Lagrangian describing the $SU(2)\times U(1)$ symmetry-breaking contains
a definite set of operators [35] whose coefficients should be determined
 experimentally.
Each specific model (e.g., ETC models, minimal SM, etc) corresponds to definite
numbers for those coefficients. One of the most important experimental tests of
a strongly-interacting Higgs sector would be the study at LHC/SSC energies of
longitudinal $W$-boson scattering, $W_LW_L\rightarrow V_LV_L$, where
$V_L=W_L, Z_L$. Since the longitudinal degrees of freedom of the massive
gauge bosons correspond to the Goldstone boson, in a theory with a
strongly interacting Higgs sector the cross section for $W_LW_L$
scattering should be large. Scatering of $W_L$s should be accesible
from $W_L$ bremstrahlung in $p-p$ collisions at LHC/SSC. Detailed computations
 [36]
show, however, that the rate for these reactions to be above background
would normally require the existence of some resonance (e.g., a
techni-$\rho $) in the $W_LW_L$ chanel.

\vskip 5.4mm
\section{ The Naturality Problem: the Supersymmetric Option}
\vskip 2.0mm
In the last twelve years Supersymmetry (SUSY) has emerged as a serious
alternative to avoid the naturality problem [37]. In this approach there are no
new strong interactions and the Higgs sector is weakly interacting.
This symmetry introduces a supersymmetric partner for each particle with
opposite statistics and spin differing by 1/2 unit. SUSY thus transforms
fermions into bosons and viceversa. The building blocks of a renormalizable
SUSY field theory are the ''chiral multiplets" and the ''vector multiplets".
 A chiral multiplet $(\psi ;\phi )$  contains
a complex scalar $\phi $ and a Weyl spinor $\psi $ whereas a vector
multiplet $(A^{\mu };\lambda )$ contains a gauge boson $A^{\mu }$ and its
''gaugino" $\lambda $ which is a Weyl spinor. The usual quark($q$), leptons
($l$) and
Higgs($H,{\bar H}$)  fields fit into chiral multiplets along with their
SUSY-partners,  the squarks(${\tilde q}$)  , sleptons($\tilde l$)
 and Higgsinos (${\tilde H},{\tilde {\bar H}}$). The $SU(3)\times SU(2)\times
U(1)$ gauge bosons fit into vector multiplets along with their
gauginos (gluinos, winos and bino). With these building blocks one easily
builds a SUSY version of the SM. Everything works as in the usual field
theory of the SM but with  additional interactions involving the
SUSY-partners. The number of coupling constants is the same as in the
non-SUSY SM. But now, due to the presence of the additional
partners and couplings, the Higgs mass parameters are stable
under radiative corrections (they are not renormalized), providing
a solution to the naturality problem.

Of course, SUSY cannot be an exact symmetry of nature and has to be
broken in some way. It is the process of SUSY breaking which introduces
additional parameters in the SUSY-SM. One may introduce terms in the Lagrangian
which explicitly break SUSY but, in order not to get the stability of the
scalar masses spoiled, these additional terms have to be of
some restricted type ({\it soft SUSY-breaking terms}). This restricted type
of terms are precisely the type of terms one obtains if SUSY
(or better, its gauge version, Supergravity) is {\it spontaneously
broken} in a ''hidden sector " of the theory, but I will not elaborate
on this point here. Let me just say that in this scheme the mass scale
of the usual SUSY partners is fixed by the mass of the ''gravitino",
the SUSY partner of the graviton. The only phenomenologically
important point is that certain SUSY-breaking soft terms appear now in
the  Lagrangian, including scalar masses, gaugino masses
and some extra scalar interactions.  In the simplest
model, the {\it Minimal Supersymmetric Standard Model} (MSSM), there are
only four SUSY-breaking soft terms:
\newpage
 i) Universal gaugino masses  { $M_{1/2}$ }

ii) Universal scalar masses  { $M_0$}

iii) Trilinear scalar couplings proportional to h${ M_0A}$

iV) Mixed Higss mass term of the type $ B\mu H{\bar H}+h.c.$.
\vskip 2.0mm
Here h is the corresponding Yukawa coupling and $\mu $ is a
possible SUSY-preserving Higgs supermultiplet mass term which may
in general be present in the original lagrangian. Thus in the MSSM
the list of new SUSY parameters is:
\vskip 2.0mm
\centerline{ $M_{1/2}$\ ,\  $M_0$\ ,\ $A$\ ,\ $B$
\ ,\ ${ \mu  }$}
\vskip 2.0mm
All these couplings are universal (e.g., all scalar masses of all different
squark, slepton and Higgs fields are equal) at a large mass scale
(the grand unification or the Planck mass scales).

One of the most atractive features of the SUSY versions of the SM is that
$SU(2)\times U(1)$ symmetry-breaking appears as a direct consequence of
SUSY-breaking. One can see that, once the above soft terms are generated,
loop effects generate a scalar potential for the Higgs fields which
automatically induces $SU(2)\times U(1)$-
breaking.
 At the unification scale all soft scalar masses are equal to
$M_0$ but the low energy evolution computed through the renormalization
group equations  is
different: the squark  $mass^2$ increases at low energies, the sleptons
and $H$ masses vary very litle whereas the $mass^2$ of the other
Higgs ${\bar H}$ becomes negative! This is precisely what we need in order
to spontaneously break $SU(2)\times U(1)$ and at the same time avoid
color or charge symmetry breaking. It is important to remark that the
above behaviour is very generic and is essentially a consequence of
the multiplet structure and quantum numbers of the SUSY-SM. One point is
however important for the mechanism to work out correctly: the
$\bar H$ Higgs field gets a negative $mass^2$ only if its coupling to the
top-quark (the t-quark Yukawa coupling) is sufficiently large.
Numerically, one essentially needs to have $m_t\geq 60$ GeV if the
mechanism is to work in a natural way. When this mechanism was
propossed these values for $m_t$ seemed fantastically large but
nowadays we know that the top-quark mass is in fact very large
and the radiative $SU(2)\times U(1)$ breaking mechanism is in fact
a very natural one. Needless to say, in order to obtain the
minimum of the Higgs potential at the correct scale (i.e. in order to
reproduce the measured values of the $W$ and $Z^0$ masses), the
full parameter space  $M_{1/2}, M_0,A,B,\mu , h_{top}$  is
constrained, but the correct numbers are obtained for very wide
ranges of the above parameters.

In the last three years or so there has been a certain increase in the
number works in the field of supersymmetry, particularly
on the minimal supersymmetric standard model. Much of this sociological
effect was motivated by the famous joining of the three gauge coupling
constants at a single unification scale which, with the advent of the
LEP data, became more striking [38]. Many of the SUSY topics studied in the
early eighties were reconsidered and analized in more detail. Some
of the SUSY topics recently reconsidered are the following:
\vskip 2.0mm
i) {\it Unification of gauge coupling constants}
\vskip 2.0mm
Indeed, when one runs up in energies [39] the three gauge coupling constants
 $g_1,g_2,g_3$
using the renormalization group, the low energy data is consistent
with unification at a single point $M_X\simeq 10^{16}$ GeV [40]. In the case of
 $g_1$,
unification takes place if the standard GUT boundary condition
 $g_1^2=3/5g_{GUT}^2$
holds. An equivalent way of stating the same fact is that, if there is
 unification
of the gauge coupling constants into a standard GUT ($SU(5)$, $SO(10)$, etc),
{\it one can compute one of the gauge couplings in terms of the others}. At the
one-loop level and ignoring threshold corrections one gets the well known
 formulae
 \begin{equation}
sin^2\theta _W(M_Z)={3\over 8}(1+{{5\alpha (M_Z)}\over {6\pi }}(b_2-{3\over
 5}b_1)
log({{M_X}\over {M_Z}}) )
\label{eq:7}
\end{equation}
\begin{equation}
{1\over {\alpha _s(M_Z)}}={3\over 8}({1\over {\alpha (M_Z)}}-{1\over {2\pi }}
(b_1+b_2-{8\over 3}b_3)log({{M_X}\over {M_Z}}))
\label{eq:8}
\end{equation}
 where in the MSSM one has $b_1=11$, $b_2=1$ and $b_3=-3$. Since the
 experimental
errors for $sin^2\theta _W$ and $\alpha $ are the smallest, what makes sense is
 to
compute $\alpha _s$ in terms of the other two.
 A number of refinements have been introduced for this
computation in the last three years including [41]: i) Effect of superheavy
 GUT-thresholds.
This can only be done in an specific GUT model like minimal SUSY $SU(5)$. ii)
Effect of the low energy sparticle thresholds.  iii)  Effect of two loop
corrections involving the t-quark.  iv) Possible corrections to the GUT values
 of
the coupling constants due to non-renormalizable effects coming (presumably)
from gravity. Including all these possible sources of uncertainties one finds
the following result [41]
\begin{eqnarray}
\alpha _s (M_Z)= 0.125 \pm  0.01
= 0.125\pm 0.001\pm 0.005{ ^{+0.005}_{-0.002}}\pm 0.005\pm 0.002\pm 0.006
\label{eq:9}
 \end{eqnarray}
The first two errors in the above expression come from the errors in the
 original
input parameters $\alpha (M_Z)$ and $sin^2\theta _W(M_Z)$. The other four
 additional
errors come from each of the four effects i)-iv) listed above. The result
 obtained
for $\alpha _s (M_Z)$ is in very good agreement with the experimental avarage
$\alpha _s (M_Z)=0.12 \pm 0.01$. It is important to remark that in the case of
 the
non-SUSY unification one obtains the prediction $\alpha _s(M_Z)=0.075$, which
is
clearly ruled out.
\vskip 2.0mm
ii) {\it $m_b/m_{\tau }$ and the mass of the top quark}
\vskip 2.0mm
In many grand unified theories like $SU(5)$ or $SO(10)$ the Yukawa couplings of
the b-quark and the $\tau $-lepton are equal at the unification scale $M_X$
 [42]. At low
energies these couplings get renormalized differently and in the SUSY case [43]
 one
then gets a predicted ratio of the type [44]:
\begin{eqnarray}
{{m_b(M_Z)}\over {m_{\tau }(M_Z)}}=({{\alpha _s(M_Z)}\over {\alpha
 _s(M_X)}})^{8/9}
 ({{\alpha _1(M_Z)}\over {\alpha _1(M_X)}})^{10/99}\
(1+{6\over {(4\pi )^2}}h^2_{top}F(M_Z))^{-1/12}
\label{eq:10}
\end{eqnarray}
 where $F(M_Z)\simeq 290$ and $h_{top}$ is the t-quark Yukawa coupling.
The first two factors come from renormalization due to gluon and hypercharge
 boson
exchange, whereas the third factor comes from loops involving a virtual top
 quark.
 Of course, in order to
compare with the physical masses of $b$ and $\tau $, one has to run $m_b$ down
 to
half the upsilon mass. In the early eighties the preferred values for the top
quark mass were relatively small and hence, the last factor in (10) was
usually ignored. Now we know that  $m_{top}$ is large and hence that term
cannot
 in
general be neglected. Notice that, if we knew $m_b$, $m_{\tau }$, $\alpha _s$
 and
$\alpha _1$ with very good precission, we should be able to extract what is the
 value
of the top Yukawa coupling [44]. With $h_{top}$ so determined, one gets a
 relationship
between $m_{top}$ and $tg\beta $ since both are related by
  \begin{equation}
m_{top}=h_{top}({{{\sqrt 2}M_W}\over {g_2}})sin\beta .
\label{eq:12}
\end{equation}
($\beta $ is defined by $tg\beta =<{\bar H}>/<H>$). There have been a number
of recent analysis of the $m_{top}$ dependence of the $m_b/m_{\tau }$ ratio
[45]-[47]. The results
are very sensitive to the value taken for $\alpha _s$ and also on the value of
$m_b$. If one takes for $\alpha _s$ the value given in eq.[9], impose the
''experimental" condition $0.85m_b^0(5 GeV)\leq 4.45 GeV$ [41] and takes the
LEP
constraints on the top mass $120 GeV\leq m_{top}\leq 160 GeV$, one finds that
 the region
$3\leq tg\beta \leq 40 $ would be forbidden. However, this bound dissapears for
 values
of $\alpha _s$ slightly smaller than $0.120$. Furthermore, slight corrections
to
 the
GUT identity $h_b=h_{\tau }$ which may come from a variety of sources close to
 the GUT scale
make also the bound on $tg\beta $ to disappear (see S. Pokorski contribution to
 the
parallel session).

One interesting point recently analized is whether a GUT boundary condition
[48] equating {\it all}
third generation Yukawa couplings $h_{top}=h_b=h_{\tau}$ is consistent both
with
data and the theoretical constraints. This type of relationship appears in
 simple
$SO(10)$ models and has the virtue of reducing the number of free parameters in
explicit models of fermion masses. The answer [41],[49] seems to be yes,
however
 one needs to
have $m_{top}\simeq 180-190$ GeV, the fixed point value. Furthermore, very
large
values for $tg\beta \simeq 50$ are required, which is very unnatural to get in
a
radiative $SU(2)\times U(1)$ breaking model [50].
\vskip 2.0mm
iii) {\it Ansatze for quark and lepton mass matrices}
\vskip 2.0mm
Another topic which has received new attention is the construction of
predictive
ansatze for fermion mass matrices within the context of SUSY-GUTs. Some simple
ansatze  with
''texture zeros" lead to attractive predictions for masses and mixing angles
like e.g., $|V_{us}|={\sqrt {m_d/m_s}}$ or $|V_{cb}|={\sqrt {m_c/m_t}}$.
The subject has been nicely reviewed in the paralell sessions by S.Raby and
G.Ross.
\vskip 2.0mm
iv) {\it SUSY-Higgs masses}
\vskip 2.0mm
This topic is of very direct phenomenological relevance [51]. The issue which
 has been
reconsidered in this case is the validity of the theoretical upper bounds on
the
lightest SUSY neutral Higgs scalar. At the {\it tree level} in the MSSM there
is
 always
a neutral scalar which is necessarily lighter than the $Z^0$ mass. On the other
 hand,
for a heavy top quark (which is the experimental case) the loop corrections to
 the
masses of the scalars  give large contributions  of order
  \begin{equation}
\delta m^2\simeq {3\over {8\pi ^2}}{{g_2^2m_{top}^4}\over {sin^2\beta M_W^2}}
log(1+{{m_{\tilde q}^2}\over {m_{top}^2}})\ .
\label{eq:12}
\end{equation}
One then finds that the lightest Higgs scalar has to be lighter than something
like $130$ GeV for values of $m_{top}$ of order $180$ GeV and squark masses
around one TeV. Thus, unfortunatelly, there is no guarantee that LEP-II will be
enough to check the Higgs sector of the MSSM.
Another important issue about the SUSY-Higgs sector is whether
the combined data from both LEP-II and LHC will be enough to probe it. If one
 draws
a plot of one the neutral Higgs mass (e.g., that of the pseudo-scalar $A$)
 versus
$tg\beta $, one finds a certain window for $5\leq tg\beta \leq 20$ and
$100 \ GeV \leq m_A \leq 200\ GeV$ in which the rate (and/or signature)
 for Higgs-particle   production is too small to be detectable. It is an
important challenge to look for interesting signatures to close this
window and some ideas on how to close it have already been put forward [52].

\vskip 2.0mm
v) {\it Other SUSY-topics}
\vskip 2.0mm
Many other areas of SUSY standard model phenomenology have been recently
 reanalized.
Amongst those the following: a) SUSY proton decay. The decay rate coming from
dimension five operators in minimal SUSY-SU(5) is close to the experimental
 limits
on nucleon instability [53]. Some authors even claim one can already constraint
the masses of SUSY-particles (or $tg\beta $) using those limits (see talks in
 the
parallel sessions by Arnowitt and Nath); b) Upper bounds on the masses of
 SUSY-particles
from naturallity arguments (see talks by Arnowitt, Nath and Ross) ; c)
 Constraints on
MSSM parameters in order to get appropriate ammount of dark matter in the form
of lightest stable neutralinos (see talks by Roszkowsky and Ross); d) SUSY
contributions to the decay $b\rightarrow s\gamma $ etc.

It is certainly intriguing how the MSSM has passed a number of important tests
 in the
last decade. I find particularly significant the joining of the three coupling
constants at a single point and also that within the MSSM a heavy top quark
 leads
in a naturall way to the spontaneous breakdown of $SU(2)\times U(1)$. The other
merits of SUSY which are ocasionally mentioned (like e.g., correct
$m_b/m_{\tau}$ por large $m_{top}$; correct ammount of dark matter predicted;
not too much proton decay etc.) are more model-dependent. One has to remark
also
that essentially all these interesting points are present not only in
the MSSM but also in simple extensions like models with R-parity
violation and models with an extra singlet Higgs scalar (sometimes
called the NMSSM =next to MSSM).

\vskip 5.4mm
\section{ Challenges for Supersymmetric Unification}
\vskip 2.0mm
 Not everything is nice and simple within the realm of the SUSY standard model.
There are a good number of issues which still need to be understood. I will
 briefly
mention here  the four problems which look more relevant to me. The first three
 are
the following: a) In the SUSY standard model the soft terms which break SUSY
are in general complex. This gives rise to new sources of CP-violation beyond
 the
KM-phase. In particular, one finds that, unless the complex phase appearing in
 the soft terms
are small (smaller than $10^{-2}-10^{-3}$), one gets large contributions to the
electric dipole moment of the neutron (EDMN) two or three orders of magnitude
 above
experimental limits; b) In SUSY versions of the SM there are also new sources
of flavour-changing neutral currents (FCNC). These are due to the exchange of
 sparticles
in box diagrams and appear if e.g. the squarks are not degenerate in mass;
c) In the models in which supersymmetry-breaking takes place in a hidden sector
 of the
theory, one can have cosmological problems with gravitino and other singlet
fields in charge of SUSY-breaking which can spoil standard nucleosintesis if
some stringent constraints on the masses of those particles are not obeyed
[54].

I must say that the above three problems are interesting constraints on
explicit models of SUSY-breaking but do not seem to me difficult to overcome.
Indeed, there are different SUSY-breaking models in the literature in which the
 above
points do not cause any trouble. On the other hand, in my opinion, the fourth
 problem that
I want to mention is really serious. This is the famous
{\it doublet-triplet splitting problem} which has been with us already for more
 than
fifteen years. Let me remark from the outset that this is {\it not } really a
 problem
of SUSY, but a problem of GUTs. Let us briefly recall what it is. Consider the
simplest case of $SU(5)$. The Weinberg-Salam doublet $H_2$ which breaks the
 electroweak
symmetry is contained in a five-plet of $SU(5)$ along with a triplet of
coloured
scalars $H_3$. The latter have to be superheavy, otherwise they would mediate
very fast proton decay through dimension six operators. Thus we have to arange
the parameters of our $SU(5)$ Lagrangian in such a way that $H_2$ remains light
 (to
be available for electro-weak symmetry breaking) but $H_3$ is superheavy (to
 avoid
fast proton decay). These doublet-triplet splitting requires a fine-tuning of
 the
Lagrangian parameters to one part in $10^{14}$!! Supersymmetry does not solve
 this
problem, it just guarantees that, provided the fine-tuning is done, radiative
corrections will not spoil it. Although some ideas in order to obtain the
 doublet
triplet splitting without fine-tuning have been suggested (sliding singlet
mechanism [55], missing partner mechanism [56], Higgses as Goldstone bosons
[57]
 etc..) all of them
either do not work (in the case of the first of them)  or are really cumbersome
 and ad-hoc.

In my opinion this doublet-triplet splitting problem of GUTs is sufficiently
 important
to consider seriously the possibility of giving up on GUTs (but not on SUSY!!).
{\it Is the GUT idea really needed?} Supersymmetry is certainly enough to solve
 the
hierarchy problem but we do not want to get rid of the nice properties of GUTs.
Amongst those one of the nicest is charge quantization, i.e. the fact that
$Q_e=3Q_d$, which is automatic in GUTs like $SU(5)$ and $SO(10)$. In fact this
charge quantization property does not require any GUT. It is well known [58]
 that
it may be equally obtained if one imposses anomaly cancellation on one family
of
quarks and leptons. Thus we really do not need unification to get that one.
The second very nice property of standard GUTs is the prediction
of gauge coupling unification which ocurrs precisely for the GUT scale
value $sin^2\theta _W=3/8$. This nice prediction is not easy to obtain in
a non-unified theory. One interesting alternative is to consider string
theories
which have the remarkable property of gauge coupling unification even in the
absence of a unification group. This bring us to our last and most speculative
subject.

\vskip 5.4mm
\section{ Superstring Phenomenology}
\vskip 2.0mm
The most  outstanding virtue of
four-dimensional (4-D) supersymmetric heterotic strings
[59] is that they are finite theories of quantum
gravity, a property which is not true of any 4-D field-theory. Furthermore,
they
allow for chiral gauge interactions like the ones of the SM. Thus 4-D strings
 are the only
known candidates for unified theory of {\it all} interactions. These are
 theories of
closed strings which contain in their spectrum an infinity of particles (string
 excitations),
all of them but a few with masses of order of the Planck mass. One identifies
 the massless
states as candidates to describe the observed world. There is a unique type of
 string
interaction (the merging of two closed strings into a single closed string) and
 this leads
to identities between the gauge coupling constants and the Newton coupling
 constant, as we
will mention below. An important property of string models is that the coupling
 constants
are not constants but fields whose vacuum expectation values determine the
 physical
values.

There are at present explicit examples of 4-D strings with three
quark-lepton generations and a gauge group containing the $SU(3)\times
 SU(2)\times U(1)$
interactions of the standard model. All these models typically have extra
 particles
(e.g., heavy leptons and/or quarks, extra gauge bosons). Although there is not
 at present
a specific
model which exactly mimics absolutely {\bf all} the desired properties of,
e.g.,
 the SUSY-SM,
they get tantalizingly close. Furthermore, these string models have some
 attractive
generic features (e.g., hierarchies of Yukawa couplings; existence of a
 multiplicity of
generations; constraints on the possible particle representations; possibility
 of
gauging some anomalous $U(1)$s etc.) which open new
avenues for model-building of unified theories. The construction of 4-D strings
 which
resemble at low energies  the SM and the study of possible generic features of
unified models based on 4-D strings goes under the name of {\it string
 phenomenology} [60].
This is a vast field and I cannot discuss all the aspects of it. I will
 concentrate
on two topics which have recently received some attention: SUSY-breaking soft
 terms
and gauge coupling unification in 4-D strings.

One of the interesting features of string models is that they have natural
 candidates
to constitue the "hidden sector" breaking supersymmetry in $N=1$ supersymmetric
 models.
In particular there are a couple of singlet scalar fields $S$ and $T$ which
 couple
to usual matter only through non-renormalizable terms supressed by inverse
 Planck mass powers [61].
The vev of these fields have a clear physical interpretation:
$<ReS>=1/g^2$ determines the size of the string (i.e. gauge) coupling constant
 $g^2$ and
$ReS$ is called the dilaton. Concerning the other one has $<ReT>=R^2$, where
$R$
 is the
overall size of the six extra compactified dimensions present in compactified
string models. The field $T$ is called the modulus. Of course it would be very
 interesting
to compute from first principles those two vevs which would tell us what should
 be the
size of the gauge couplings $g^2$ and what is the size of the compactification
 radius $R^2$.
The latter should be of order one in Planck mass units. To find those vevs we
 would need
to know the scalar potential $V(S,T)$ and then look for its minima.
 Unfortunately
this potential vanishes order by order in perturbation theory and hence $ReS$,
 $ReT$
are undetermined at this level. Thus the expectation is that non-perturbative
effects will raise that degeneracy and create a non-vanishing potential for $S$
 and $T$.
Indeed in simple models of "gaugino condensation" [62] such non-vanishing
 potentials are in fact
generated. In these models
it is assumed that the gaugino fields $\lambda$  corresponding to  extra
 ("hidden sector") gauge
factors normally present in string models condense (i.e. $<\lambda \lambda
 >\not=0$) and this gives
rise to SUSY-breaking and a non-vanishing scalar potential $V(S,T)$
[63],[64]. For particular classes
of hidden-sector gauge groups and particle content, the values obtained for
 $<ReS>=1/g^2$
are consistent with the extrapolation of the low energy gauge coupling
constants
 [65]. Within
this type of models one can also obtain results for the values of the
 SUSY-breaking soft
terms $M_{1/2},M_0,A,B$ that we discussed above [66].

There is some more general information about soft terms which can be obtained
 without
resorting to specific gaugino condensation models.
One can find simple expressions for the SUSY-breaking soft terms if one assumes
 that
the two fields $S,T$ play the leading role in providing the source for
 SUSY-breaking
[67]-[70].
One does not need to know the precise way in which SUSY-breaking takes place to
 get those
results. In this case the "goldstino" $\tilde {\eta } $ (the Goldstone particle
 of SUSY-breaking) is
predominantly a linear combination ${\tilde {\eta }}=sin\theta {\tilde
 S}+cos\theta {\tilde
T}$ of the fermionic partners of $S$ and $T$. For $sin\theta =1$ the dilaton
$S$
 is the
leading source of SUSY-breaking whereas for $cos\theta =1$ it is the modulus
$T$
 which is
dominant. Now, depening on the particular string model and the value of
 $sin\theta $ one
gets different results for the soft terms [70]. The effective low energy
 Lagrangian of the model
is determined by the standard $N=1$ supergravity functions which are the Kahler
potential $K(S,T,C_i)$ and the gauge kinetic function $f^a$, $a$ being a gauge
 group label.
The tree level form of $f^a$ is universal for any string model, $f^a=k^aS$,
 where the $k^a$
are numerical constants to be discussed below. The kahler potential $K$ is a
 model-dependent
function. In a simple (but large) class of models (symmetric orbifolds) this
 function
takes the general form [71]:
  \begin{eqnarray}
K(S,T,C_i)\ =\ -log(S+S^*)\ -\ 3log(T+T^*)\ + \
  +(T+T^*)^{n_i}C_iC^*_i
\label{eq:13}
\end{eqnarray}
where the $C_i$ correspond to the matter fields, like
(s)quarks, (s)leptons etc. The $n_i$
are particle-dependent integers whose most common values are -1,-2,-3 for this
 class of
models [68]. One can plugg this expression in the general form of the scalar
 potential and assume
that $S$ and $T$ give the dominant source of SUSY-breaking. If one farther
 imposses that the
cosmological contant should vanish [69], one finds expressions for the soft
 terms of the
following type [70]:
 \begin{eqnarray}
m_i^2= m_{3/2}^2(1+n_icos^2\theta )  \ \ \ \ \ \ \ \ \ .\nonumber \\
 M_{1/2}={\sqrt 3} m_{3/2}sin\theta \ \ \ \ \ \ \ \ \ \ \ . \nonumber \\
 A^0_{ijk} = -{\sqrt 3} m_{3/2}(sin\theta +
{{(3+n_i+n_j+n_k)}\over {\sqrt 3}}cos\theta )
\label{eq:14}
\end{eqnarray}
where the indices $ijk$ refer to the three particles coupled through the
 $A$-term
and the superindex $0$ indicates that this is the $A$ parameter of a large,
 non-supressed
Yukawa coupling (like the one expected for the top). In some
classes of orbifolds (e.g. $Z_2\times Z_2$) and for the large $T$ limit of
 Calabi Yau
compactifications one has $n_i=-1$ and then one gets the even simpler boundary
condition [69],[70]:
   \begin{equation}
M_{1/2}\ =\ -A\ =\ {\sqrt 3}\ m_i\ \ .
\label{eq:15}
 \end{equation}
Notice that in fact those boundary conditions also apply in the limit
$sin\theta
 =1$
(dilaton dominance limit) for any $n_i$ choice, i.e., in a model-independent
way
 [69],
since the dilaton dependence of the effective low energy Lagrangian is indeed
model-independent. Although the above expressions for the soft terms involve
some assumptions (most notably that $S,T$ dominate SUSY-breaking and the
impossition that the cosmological constant vanishes) it is remarkable how far
 one
can go in obtaining equations based on string physics which may be
 experimentally tested.
Indeed, expressions like the ones above lead to certain constraints on the low
 energy
supersymmetric mass spectrum when run down in energies according to the
 renormalization
group equations. Such type of analysis has been recently done in ref.[72],[70].

Another interesting topic in string phenomenology is the issue of gauge
coupling
unification. Suposse we had a 4-D string with gauge group including the one of
the standard model. Then one finds [73]
   \begin{equation}
g_3^2k_3\ =\ g_2^2k_2\ =\ g_1^2k_1\ =\ {{4\pi M_{string}^2}\over
{M_{Planck}^2}}
\label{eq:16}
 \end{equation}
Thus {\it even in the absence of a GUT group} there is unification of coupling
constants. The $k_a$ corresponding to the non-Abelian gauge factors are
positive
integers called the Kac-Moody levels. In practically all models considered up
to
 now
one has $k_2=k_3=1$. In fact this is always the case for string models obtained
upon direct compactification of the $E_8\times E_8$ heterotic string. The
 corresponding
constant $k_1$ for abelian factors like the hypercharge  is a model-dependent
rational number. Consistency with the low energy spectrum of the standard model
requires $k_1\geq 1$ [74] but otherwise there is no other model independent
 constraint
on it. This leads to the model-independent constraint
$sin^2\theta _W(M_{string})=k_2/(k_1+k_2)=1/(1+k_1)\leq 1/2$.  The standard GUT
value for $k_1$ is $5/3$, leading to $sin^2\theta _W(M_{GUT})=3/8$, but in the
non-unified string case $k_1$ needs not be equall to the GUT result [75].
 Since the couplings are unified at the
string scale $M_{string}\simeq 4\times 10^{17}$ GeV [76] and not at the
 previously
mentioned SUSY-GUT scale $M_X$, instead of eqs.[7,8] one gets:
 \begin{eqnarray}
 sin^2\theta _W(M_Z)= \
{1\over {1+k_1}}(1+ {{k_1\alpha (M_Z)}\over {2\pi }}(b_2-{{b_1}\over {k_1}})
log({{M_{string}}\over {M_Z}}) )
\label{eq:17}
\end{eqnarray}
\begin{eqnarray}
 {1\over {\alpha _s(M_Z)}}=  \
{1\over {1+k_1}}({1\over {\alpha (M_Z)}}-{1\over {2\pi }}
(b_1+b_2-(1+k_1)b_3)
log({{M_{string}}\over {M_Z}}))
\label{eq:18}
\end{eqnarray}
Let us assume for a moment that a) $k_1=5/3$ as happens e.g. in $E_6$ type of
 string
compactifications; b)  that the only particles charged under
$SU(3)\times SU(2)\times U(1)$ are the ones of the minimal supersymmetric
 standard model
and c) neglect possible string threshold effects. Then one finds
[77],[78] $sin^2\theta _W(M_Z)=0.218$ and
$\alpha _s(M_Z)=0.20$, far away from the experimental results (but still better
 than non-SUSY
$SU(5)$). Thus one has to give up at least one of the above three assumptions
a)
 to c).
One can give up assumption b) and consider further charged particles
apart from those of the MSSM. If the extra introduced particles are not to be
 exotic
this will require the introduction of some intermediate mass scale(s) between
 $M_{Planck}$
and the weak scale [78]. This opens a Pandora's box of possibilities.
Concerning
 assumption
c), indeed string threshold effects can be relatively large, since infinite
 towers of
particles may give contributions in the vecinity of $M_{string}$. String
 threshold
effects have been computed in some class of orbifold models [79].
 Phenomenological analysis
shows [77],[68] that the string threshold corrections may be large as long as
 $ReT\simeq 4-20$.
These corrections go in the good direction only for very restricted possible
 models.
Finally, one can give up assumption a) and consider values of $k_1$ different
 from
$5/3$ [75]. The best agreement is found for $4/3\leq k_1\leq 3/2$. For
 $k_1=1.444$ one finds
$sin^2\theta _W(M_Z)=0.233$ and $\alpha _s(M_Z)=0.14$ ; for $k_1=1.466$ one
gets
$sin^2\theta _W(M_Z)=0.235$ and $\alpha _s(M_Z)=0.137$. Given  the inherent
uncertainties coming from string threshold effects these results can be
 considered
as succesfull, and show that claiming for the necessity of intermediate scales
 in direct string
unification may be premature.
On the other hand specific string models with $k_1$  in the interesting range
still have to be found. In any case it is
clear that the constraint of gauge coupling unification in string models may be
 very important in
selecting  possible string unification schemes.

 \vskip 5.4mm
\section{ Conclusion}
\vskip 2.0mm

The general field of Particle Theory beyond the Standard Model is quite
 speculative and,
at the moment, it
is quite difficult to favour on a firm basis any of the avenues discussed above
 better than
any other. It may well be that the directions of the arrows showing the
tendency
 of
different topics  listed in the introduction flip in some cases upside down.
 Our main
hope relies on the planned new experiments and it is also our hope that some of
 the
"trends" briefly described above will be promoted to real physics!

\vskip 5.4mm
\newpage
\vskip 2.0mm

\centerline{\bf  REFERENCES}
\vskip 2.0mm

\noindent
[1] G. t'Hooft,
{\it Phys.Rev.Lett.} {\bf 37}
(1976) 8.
\newline
\noindent
[2] V. Kuzmin, V. Rubakov and M. Shaposhnikov, ,
{\it Phys.Lett.} {\bf B155}
(1985) 36;P. Arnold and L. McLerran, {\it Phys.Rev.} {\bf D36} (1987) 581;{\bf
 D37}(1988) 1020.
\newline
\noindent
[3] F. Klinkhamer and N. Manton,
{\it Phys.Rev.} {\bf D30}
(1984) 2212.
\newline
\noindent
[4] For a review and references see: M. Shaposhnikov, CERN-TH.6497/92; M. Dine,
 Santa Cruz
preprint
 SCIPP 92/21 (1992); A. Cohen, D. Kaplan and A. Nelson, UCSD-PTH-93-02,
 BUHEP-93-4
(1993).
\newline
\noindent
[5] A.D. Sakharov, {\it Pisma ZhETF} {\bf 5} (1967) 32; V.A. Kuzmin, {\it Pisma
 ZhETF}
{\bf 13} (1970) 35; L. Okun and Y. Zeldovich, {\it Comm. Nucl. Part. Phys.}
{\bf
 6}
(1976) 69.
\newline
\noindent
[6] G. Farrar and M. Shaposhnikov, {\it Phys.Rev.Lett} {\bf 70} (1993) 2833;
 CERN-TH-6732-93;
M.B. Gavela, P. Hernandez, J. Orloff and O. Pene, CERN-TH-7081-93
\newline
(1993).
\newline
\noindent
[7] A.Y. Ignatiev, N. Krasnikov, V. Kuzmin and A. Takhelidze, {\it Phys.Lett.}
 {\bf 76B}
(1978) 436; M. Yoshimura, {\it Phys.Rev.Lett.} {\bf 41} (1978) 4500;
S. Dimopoulos and L. Susskind, {\it Phys. Rev.} {\bf D18} (1978) 4500;
S. Weinberg, {\it Phys.Rev.Lett.} {\bf 42}  (1979) 850.
\newline
\noindent
[8] See e.g. M.A. Luty, {\it Phys.Rev.} {\bf D45} (1992) 455  and references
 therein.
\newline
\noindent
[9] M. Fukugita and T. Yanagida, {\it Phys.Rev.} {\bf D42} (1990) 1285;
J. Harvey and M. Turner, {\it Phys.Rev.} {\bf D42} (1990) 3344;
A. Nelson and S. Barr, {\it Phys.Lett.} {\bf B246} (1991) 141.
\newline
\noindent
[10] B. Campbell, S. Davidson, J. Ellis nd K. Olive, {\it Phys.Lett.} {\bf
B256}
 (1991) 457;
W. Fischler, G. Giudice, R. Leigh and S. Paban, {\it Phys.Lett.} {\bf B258}
 (1991) 45.
\newline
\noindent
[11] L.E. Ib\'a\~nez and F. Quevedo, {\it Phys.Lett.} {\bf B283} 261.
\newline
\noindent
[12] I. Affleck and M. Dine, {\it Nucl.Phys.} {\bf B249} (1985) 361.
\newline
\noindent
[13] L.E. Ib\'a\~nez and F. Quevedo, Proc. of the Texas/Pascos Conf., Berkeley,
 Dec. 1992.
World Scientific eds.
\newline
\noindent
[14] See e.g. T. Banks, {\it Physicalia Magazine} {\bf 12} (1990) 19;
G. Gilbert, {\it Nucl.Phys.} {\bf B328} (1989) 159.
\newline
\noindent
[15] R. Peccei and H. Quinn, {\it Phys.Rev.Lett.} {\bf 38} (1977) 1440;
{\it Phys.Rev.} {\bf D16} (1977) 1791;
S. Weinberg,
{\it Phys.Rev.Lett.} {\bf 40 } (1978) 223;
F. Wilczek, {\it Phys.Rev.Lett.} {\bf 40} (1978) 279.
\newline
\noindent
[16] M. Kamionkowski and J. March-Russell,
{\it Phys.Lett.} {\bf B282} (1992) 137;
R. Holman et al. {\it Phys.Lett.} {\bf B282} (1992) 132;
S. Barr, and D. Seckel, {\it Phys.Rev.} {\bf D46} (1992) 539.
\newline
\noindent
[17] A. Nelson, {\it Phys.Lett.} {\bf B136} (1984) 387;
S. Barr, {\it Phys.Rev.Lett.} {\bf 53 }  (1984) 329.
\newline
\noindent
[18] H. Georgi and I.N. McArthur, Harvard Report HUTP-81/A001 (1981),
 unpublished;
D. Kaplan and A. Manohar, {\it Phys.Rev.Lett.} {\bf 56}  (1986) 2004.
\newline
\noindent
[19] G. Lazarides, C. Panagiotakopoulos and Q. Shafi,
{\it Phys.Rev.Lett.} {\bf 56} (1986) 432;
J.A. Casas and G.G. Ross, {\it Phys.Lett.} {\bf B192} (1987) 119.
\newline
\noindent
[20] J.E. Kim, {\it Phys.Lett.} {\bf B207} (1988) 434;
L.E. Ib\'a\~nez, {\it Phys.Lett.} {\bf B303}  (1993) 55.
\newline
\noindent
[21] L. Krauss and F. Wilczek, {\it Phys.Rev.Lett.} {\bf 62} (1989) 1221;
J. Preskill, {\it Ann.Phys.} {\bf 210} (1991) 323.
\newline
\noindent
[22] T. Banks, {\it Nucl.Phys.} {\bf B323} (1989) 90;
M. Alford, J. March-Russell and F. Wilczek, {\it Nucl.Phys.} {\bf B337} (1990)
 695;
J. Preskill and L. Krauss, {\it Nucl.Phys.} {\bf B341} (1990) 50;
M. Alford, S. Coleman and J. March-Russell, {\it Nucl.Phys.} {\bf B351} (1991)
 735.
\newline
\noindent
[23] L.E. Ib\'a\~nez and G.G. Ross, {\it Phys.Lett.} {\bf B260} (1991) 291;
{\it Nucl.Phys.} {\bf B368} (1992) 3.
\newline
\noindent
[24] T. Banks and L. Dixon, {\it Nucl.Phys.} {\bf B307} (1988) 93.
\newline
\noindent
[25] M. Dine, R. Leigh and D. MacIntire, {\it Phys.Rev.Lett.} {\bf 69} (1992)
 2030;
K.W. Choi, D. Kaplan and A. Nelson, {\it Nucl.Phys.} {\bf B391} (1993) 515.
\newline
\noindent
[26] T. Banks and M. Dine, {\it Phys.Rev.} {\bf D45 } (1992) 1424;
L.E. Ib\'a\~nez, {\it Nucl.Phys.} {\bf B398} (1993) 301;
L.E. Ib\'a\~nez and D. L\"ust, {\it Phys.Lett.} {\bf B302} (1993) 38;
G.G. Ross and C. Scheich, {\it Phys.Lett} {\bf B315} (1993) 88;
C. Scheich, Oxford preprint OUTP-93-15P (1993);
D. MacIntire, SCIPP-93-27 (1993).
\newline
\noindent
[27] S. Weinberg, {\it Phys.Rev.} {\bf D13}  (1976)  974; {\bf D14}  (1979)
 1277;
L. Susskind, {\it Phys.Rev.} {\bf D20}  (1979) 2619.
\newline
\noindent
[28] S. Dimopoulos and L. Susskind, {\it Nucl.Phys.} {\bf B155} (1979) 237;
E. Eichten and K. Lane, {\it Phys.Lett.} {\bf B90} (1980) 125.
\newline
\noindent
[29] For a review see M.B. Einhorn in {\it Perspectives on Higgs Physics}, G.
 Kane
ed., World Scientific (1993).
\newline
\noindent
[30] T. Appelquist, D. Karabali and L. Wijewardhana, {\it Phys.Rev.Lett.}
{\bf 57} (1986) 957; T. Appelquist and L. Wijewardhana, {\it Phys.Rev.}
{\bf D35} (1987) 774; {\bf D36} (1987) 568.
\newline
\noindent
[31] B. Holdom, {\it Phys.Rev.} {\bf D24}  (1981)  1441.
\newline
\noindent
[32] G. Giudice and S. Raby, {\it Nucl.Phys.} {\bf B368}  (1992) 221.
\newline
\noindent
[33] For a review see C.T. Hill in {\it Perspectives on Higgs Physics}, G. Kane
ed., World Scientific (1993).
\newline
\noindent
[34] See e.g. H. Leutwyler and H. Georgi in the Proc. of the 1991 TASI School
 (Boulder, Co.).
\newline
\noindent
[35] T. Appelquist and C. Bernard, {\it Phys.Rev.} {\bf  D22} (1980) 200;
A. Longhitano, {\it Nucl.Phys.} {\bf B188} 118;
M. Chanowitz, H. Georgi and M. Golden, {\it Phys.Rev.} {\bf D36} (1987) 1490.
\newline
\noindent
[36] A. Dobado, M.J. Herrero and J. Terron, Proc. of the Aachen ECFA Workshop,
 p.768 (1990);
A. Dobado and M.J. Herrero, {\it Z.Phys.} {\bf C50} (1991) 205;
J. Donaghue, C. Ramirez and G. Valencia, {\it Phys.Rev.} {\bf D39}  (1989)
1947;
J. Bagger, S. Dawson and G. Valencia, Fermilab-pub
-92-75-T (1992).
\newline
\noindent
[37] For  recent phenomenological reviews see:
L.E. Ib\'a\~nez and G.G. Ross in {\it Perspectives on Higgs Physics}, G. Kane
 editor, World
Scientific (1993); F. Zwirner, CERN-TH -6951-93 (1993); H. Haber, SCIPP-99-33
 (1993).
\newline
\noindent
[38] J. Ellis, S. Kelley and D.V. Nanopoulos, {\it Phys.Lett.} {\bf B249}
(1990)
  441;
{\bf B260} (1991) 131; P. Langacker and M. Luo, {\it Phys.Rev.} {\bf D44}
(1991)
 817;
U. Amaldi, W. de Boer and H. F\"urstenu, {\it Phys.Lett.} {\bf B260} (1991)
447.
\newline
\noindent
[39] H. Georgi, H. Quinn and S. Weinberg, {\it Phys.Rev.Lett.} {\bf 33} (1974)
 451.
\newline
\noindent
[40] S. Dimopoulos, S. Raby and F. Wilczek {\it Phys.Rev.} {\bf D24}  (1981)
 1681;
S. Dimopoulos and H. Georgi,
{\it Nucl.Phys.} {\bf B193} (1982) 475;
L.E. Ib\'a\~nez and G.G. Ross, {\it Phys.Lett.} {\bf B105} 439.
\newline
\noindent
[41] For a detailed study and references see e.g.:
P. Langacker and N. Polonsky, {\it Phys.Rev.} {\bf D47}  (1993) 4028.
\newline
\noindent
[42] M. Chanowitz, J. Ellis and M.K. Gaillard, {\it Nucl.Phys.} {\bf B135}
 (1978) 66;
A. Buras, J. Ellis, M.K. Gaillard and D.V. Nanopoulos, {\it Nucl.Phys.} {\bf
 B195}
(1978) 66.
\newline
\noindent
[43] M. Einhorn and D.R.T. Jones, {\it Nucl.Phys.} {\bf B196} (1982) 475;
K. Inoue et al. {\it Prog. Theor. Phys.} {\bf 67} (1982) 1859.
\newline
\noindent
[44] L.E. Ib\'a\~nez and C. L\'opez, {\it Nucl.Phys.} {\bf B233} (1984) 511.
\newline
\noindent
[45] H. Arason et al., {\it Phys.Rev.Lett.} {\bf 67} (1991) 2933;
S.Kelley, J. L\'opez and D.V. Nanopoulos, {\it Phys.Lett.} {\bf B278} (1992)
 140.
\newline
\noindent
[46] S. Dimopoulos, L. Hall and S. Raby, {\it Phys.Rev.Lett.} {\bf 68}  (1992)
 1984;
{\it Phys.Rev.} {\bf D45} (1992) 4192.
\newline
\noindent
[47] V. Barger, M.S. Berger and P. Ohman, {\it Phys.Rev.} {\bf D47} (1993)
1093;
M. Carena, S. Pokorski and C.E.M. Wagner, {\it Nucl.Phys.} {\bf B406} (1993)
59;
P. Langacker and N. Polonsky, preprint UPR-0556T, (1993).
\newline
\noindent
[48] B. Ananthanarayan, G. Lazarides and Q. Shafi,
{\it Phys.Rev.} {\bf D44} (1991) 1613.
\newline
\noindent
[49] P. langacker and N. Polonsky, preprint UPR-0556T, (1993);
L.J. Hall, R. Rattazzi and U. Sarid,
LBL-33997 (1993). \newline
\noindent
[50] A. Nelson and L. Randall, UCSD-PTH-93-24 \newline  (1993).

\noindent
[51] For a review and references see F. Zwirner, CERN-TH-
6792-93 (1993).
\newline
\noindent
[52] See e.g. J. Dai, J. Gunion and R. Vega, SLAC-PUB-6277 (1993)
and references therein.
\newline
\noindent
[53] P. Nath and R. Arnowitt, {\it Phys.Rev.Lett.} {\bf 70} (1993) 3696;
J. Hisano, H. Murayama and T. Yanagida,
{\it Nucl.Phys.} {\bf B402} (1993) 46.
\newline
\noindent
[54] B. de Carlos, J.A. Casas, F. quevedo and E. Roulet, CERN-TH.6958/93
(1993);
T. Banks, D. Kaplan and A. Nelson, UCSD/PTH 93-26 (1993).
\newline
\noindent
[55] E. Witten, {\it Phys.Lett.} {\bf B105} (1981) 267;
L.E. Ib\'a\~nez and G.G. Ross {\it Phys.Lett.} {\bf B110} (1982) 227.
\newline
\noindent
[56] S. Dimopoulos and F. Wilczek, Santa Barbara preprint (1981);
Proc. Erice Summer School (1981);
B. Grinstein, {\it Nucl.Phys.} {\bf B206} (1982) 387;
A. Masiero, D.V. Nanopoulos, K. Tamvakis and T. Yanagida,
{\it Phys.Lett.} {\bf B115}  (1982) 380.
\newline
\noindent
[57] K. Inoue, A. Kakuto and T. Takano, {\it Prog.Theor.} \newline
{\it Phys.} {\bf 75} (1986) 664;
A. Anselm and A. Johansen, {\it Phys.Lett.} {\bf B200} (1988) 331;
R. Barbieri, G. Dvali and A. Sturmia, {\it Nucl.Phys.} {\bf B391} (1993) 487;
R. Barbieri, G. Dvali and M. Moretti, CERN-TH-6840/93 (1993).
\newline
\noindent
[58] See e.g. L.E. Ib\'a\~nez in Proc. of the V-th ASI on Techniques and
 Concepts of
High Energy Physics (St. Croix, Virgin Islands), July 1988,ed. T. Ferbel
 (Plenum, New York,
1989).
\newline
\noindent
[59]  For a collection of some relevant papers see: {\it String theory in four
 dimensions},
ed. M. Dine, North-Holland (1988);
{\it Superstring construction}, ed. B. Schellekens, North-Holland (1989).
\newline
\noindent
[60] For more phenomenological reviews see e.g.:
L.E. Ib\'a\~nez in Proc. of the XVIII International GIFT Seminar on Theoretical
 Physics,
El Escorial 1987, World Scientific (1988), CERN-TH.4769/87;
{\it Topics in String Unification} in Proc. of the Workshop on electroweak
 physics
beyond the standard model, Valencia, 1991, World Scientific (1992);
M. Dine, {\it Topics in String Phenomenology}, Proc. of String 93 (Berkeley,
May
 1993),
preprint SCIPP-93-30;
D. L\"ust, CERN-TH.6819-93.
\newline
\noindent
[61] E. Witten, {\it Phys.Lett.} {\bf B155} (1985) 151.
 \newline
\noindent
[62] H.P. Nilles, {\it Phys.Lett.} {\bf B115} (1982) 193;
S. Ferrara, L. Girardello and H.P. Nilles, {\it Phys.Lett. } {\bf B125} (1983)
 457.
\newline
\noindent
[63] J.P. Derendinger, L.E. Ib\'a\~nez and H.P. Nilles, {\it Phys.}
{\it Lett.} {\bf B155} (1985) 65;
M. Dine, R. Rohm, N. Seiberg and E. Witten, {\it Phys.Lett.} {\bf  B156} (1985)
 55;
N. Krasnikov, {\it Phys.Lett.} {\bf  B193} (1987) 37;
L. Dixon in Proc. of the APS DPFmeeting, Houston, (1990);
J.A. Casas, Z. Lalak, C. Mu\~noz and G.G. Ross, {\it Nucl.Phys.} {\bf B347}
 (1990) 243.
\newline
\noindent
[64] A. Font, L.E. Ib\'a\~nez, D. L\"ust and F. Quevedo, {\it Phys.Lett. }
{\bf B245} (1990) 401;
S. Ferrara, N. Magnoli, T. Taylor and G. Veneziano, {\it Phys.Lett.} {\bf B245}
 (1990) 409;
H.P. Nilles and M. Olechovsky, {\it Phys.Lett.} {\bf B248} (1990) 268;
P. Binetruy and M.K. Gaillard, {\it Phys.Lett.} {\bf B253} (1991) 119.
\newline
\noindent
[65] B. de Carlos, J.A. Casas and C. Mu\~noz, {\it Nucl.Phys.} {\bf B399}
(1993)
 623.
\newline
\noindent
[66] B. de Carlos, J.A. Casas and C. Mu\~noz, {\it Phys.Lett.} {\bf B299}
(1993)
 234.
\newline
\noindent
[67] M. Cvetic, A. Font, L.E. Ib\'a\~nez, D. L\"ust and F. Quevedo, {\it
 Nucl.Phys.} {\bf B361}
(1991) 194.
\newline
\noindent
[68] L.E. Ib\'a\~nez and D. L\"ust, {\it Nucl.Phys.} {\bf B382} (1992) 305.
\newline
\noindent
[69] V. Kaplunovsky and J. Louis, {\it Phys.Lett.} {\bf B306} (1993) 269.
\newline
\noindent
[70] A. Brignole, L.E. Ib\'a\~nez and C. Mu\~noz, Madrid preprint FTUAM-26/93
 (1993).
\newline
\noindent
[71] L. Dixon, V. Kaplunovsky and J. Louis, {\it Nucl.Phys.} {\bf B329} (1990)
 27.
\newline
\noindent
[72] R. Barbieri, J. Louis and m. Moretti, {\it Phys.Lett.} {\bf B312} (1993)
 451;
J. L\'opez, D.V. Nanopoulos and A. Zichichi, CERN-TH-6926-93 (1993).
\newline
\noindent
[73] P. Ginsparg, {\it Phys.Lett.} {\bf B197} (1987) 139.
\newline
\noindent
[74] A.N. Schellekens, {\it Phys.Lett.} {\bf B237} (1990) 363.
\newline
\noindent
[75] L.E. Ib\'a\~nez, {\it Phys.Lett.} {\bf B318} (1993) 73.
\newline
\noindent
[76] V. Kaplunovsky, {\it Nucl.Phys.} {\bf B307} (1988) 145 and erratum (1992).
\newline
\noindent
[77] L.E. Ib\'a\~nez, D. L\"ust and G.G. Ross, {\it Phys.Lett.} {\bf B272}
 (1991) 251.
\newline
\noindent
[78] I. Antoniadis, J. Ellis, S. Kelley and D.V. Nanopoulos, {\it Phys.Lett.}
 {\bf B271}
(1991) 31.
\newline
\noindent
[79] L. Dixon, V. Kaplunovsky and J. Louis, {\it Nucl.Phys.} {\bf B355}  (1991)
 649;
J.P. Derendinger, S. Ferrara, C. Kounnas and F. Zwirner, {\it Nucl.Phys.} {\bf
 B372} (1992) 145;
I. Antoniadis, K. Narain and T. Taylor, {\it Phys.Lett.} {\bf B267} (1991) 37.
\newline
\noindent
[80] J. Cline, K. Kainulainen and K. Olive, preprint UMN-TH-1201/93 (1993); A.
 Antaramian,
L.J. Hall and A. Ravsin, preprint LBL-34827 (1993).

\end{document}